\documentclass[12pt]{article}
\newcommand{\bd}{\begin{displaymath}}
\newcommand{\ed}{\end{displaymath}}
\newcommand{\be}{\begin{equation}}
\newcommand{\ee}{\end{equation}}
\newcommand{\bq}{\begin{quote}}
\newcommand{\eq}{\end{quote}}
\newcommand{\ben}{\begin{enumerate}}
\newcommand{\een}{\end{enumerate}}
\newcommand{\bi}{\begin{itemize}}
\newcommand{\ei}{\end{itemize}}
\newcommand{\bdes}{\begin{description}}
\newcommand{\edes}{\end{description}}

\begin{document}
\setlength{\baselineskip}{16pt}
\title{MAKING SENSE OF A WORLD OF CLICKS}
\author{Ulrich Mohrhoff\\
Sri Aurobindo International Centre of Education\\
Pondicherry 605002 India\\
\normalsize\tt ujm@satyam.net.in}
\date{}
\maketitle
\begin{abstract}
\normalsize\noindent
In a recent article O. Ulfbeck and A. Bohr (Foundations of Physics 31, 757, 2001) have 
stressed the genuine fortuitousness of detector clicks, which has also been pointed out, in 
different terms, by the present author (American Journal of Physics 68, 728, 2000). In spite 
of this basic agreement, the present article raises objections to the presuppositions and 
conclusions of Ulfbeck and Bohr, in particular their rejection of the terminology of indefinite 
variables, their identification of reality with ``the world of experience,'' their identification of 
experience with what takes place ``on the spacetime scene,'' and the claim that their 
interpretation of quantum mechanics is ``entirely liberated'' from classical notions. An 
alternative way of making sense of a world of uncaused clicks is presented. This does not 
invoke experience but deals with a free-standing reality, is not fettered by classical 
conceptions of space and time but introduces adequate ways of thinking about the 
spatiotemporal aspects of the quantum world, and does not reject indefinite variables but 
clarifies the implications of their existence.
\setlength{\baselineskip}{14pt}
\end{abstract}

\section{\large INTRODUCTION}

In a recent article O. Ulfbeck and A. Bohr~\cite{UlfBohr} have pointed out certain 
features of quantum mechanics (QM) the significance of which has not been sufficiently 
appreciated. Foremost among these is the {\it genuine fortuitousness\/} of detector clicks, 
which are usually thought of as being caused (``triggered'') by the impact of a particle: 
``the individual fortuitous event, which quantum mechanics deals with, comes by itself, 
without any cause, and is entirely beyond theoretical analysis.'' (Quotations without 
reference numbers are from Ref.~1.)

The idea that a detector click is ``a totally lawless event'' is not entirely unheard-of. In a 
letter to Max Born, Wolfgang Pauli spoke of the ``appearance of a definite position $x_0$ 
during an observation\dots as a creation existing outside the laws of 
nature"~\cite{EinBorn}. More recently the present author has argued that a 
value-indicating event such as the click of a detector is what philosophers call a {\it causal 
primary\/}---it occurs without a cause and is therefore fundamentally 
inexplicable~\cite{Mohrhoff00}.

According to Ulfbeck and Bohr, QM is exclusively concerned with distributions of clicks: 
``The wave function enters in the sole role of encoding the probability distributions of 
clicks.'' This view is close to the ``correlation interpretation''~\cite{Laloe}, a common 
ground that ought to be acceptable to everyone, inasmuch as within its minimal 
conceptual framework ``no conflict of postulates takes place''~\cite{Laloe}. What divides 
the physics community is the attempt to go beyond the correlations and say something 
reasonable about the correlata and their relation to the correlations. I fully agree with 
Ulfbeck and Bohr that such an attempt ought to take due account of the genuine 
fortuitousness of the correlata. This constraint, however, leaves plenty of room for 
disagreement, as will become obvious in what follows.

\section{\large WHY FORTUITOUS?}

My first dissent concerns the reason why the correlata are uncaused. Ulfbeck and Bohr 
aim to establish the genuine fortuitousness of detector clicks by pointing to the immense 
complexity and consequent uniqueness of each ``macroclick,'' when this is seen in 
sufficiently high resolution: ``the click involves such an immense number of degrees of 
freedom that two clicks are never identical\dots. Laws for clicks arise when the click\dots 
is seen with a lowered resolution\dots. With the lowered resolution, clicks can repeat 
thereby defining probability distributions that are within the domain of law.'' What is 
actually addressed here is the measurability of probabilities in terms of relative 
frequencies. The probability of a unique event cannot be measured. QM, however, assigns 
probabilities regardless of whether they can be measured.

That the above argument is irrelevant to the genuine fortuitousness of detector clicks is 
also obvious from the authors' independent insistence that the onset of a click, ``from 
which the click evolves as a signal in the counter\dots, does not belong to a chain of causal 
events'' and ``is not the effect of something.'' As I see it, the reason for this is that the 
probability that a variable $Q$ has the value $v$ is the product of two probabilities---the 
probability that any one of the possible values of $Q$ is indicated, and the probability 
that the indicated value is~$v$ given that a value is indicated. QM is exclusively concerned 
with probabilities of the latter type. It does not assign a probability to the occurrence of a 
value-indicating event, nor does it specify sufficient conditions for such an event. If QM is 
a fundamental and universal theoretical framework, this means that the value-indicating 
events presupposed by QM are uncaused.

\section{\large INDEFINITENESS}
\label{Indef}
My second and perhaps chief dissent concerns the authors' rejection of the terminology of 
indeterminate variables. There is no doubt that this terminology is potentially confusing, 
but given the obvious importance of the indefiniteness or relative positions for the 
existence of spatially extended material objects~\cite{Lieb}, what is needed is not a 
wholesale rejection but a conceptual clarification.

The proper way of dealing with indefinite values is to make counterfactual probability 
assignments~\cite{Mohrhoff00,Mohrhoff01}. If we say that a variable has an ``indefinite 
value,'' what we mean is that it does not have a value (inasmuch as no value is indicated) 
but that it {\it would\/} have a value if one {\it were\/} indicated, and that positive 
probabilities are associated with at least two possible values. (While the reference to 
counterfactuality cannot be eliminated, it may be shifted from values that are only 
counterfactually indicated to values that are only counterfactually indefinite: If a certain 
measurement is performed on an ensemble of identically ``prepared'' systems and the 
results exhibit a positive dispersion, the value of the measured variable would be 
indefinite for each system if the measurement were not performed.)

If a variable sometimes does and sometimes does not have a value, a criterion is called for, 
and this is the existence of a value-indicating fact. In other words, the matrix variables of 
QM are {\it extrinsic\/}, in the sense that each possesses a value only if, and to the 
extent that, a value is indicated (by an actual event or state of affairs).

Ulfbeck and Bohr object to such statements as the following, which they find ``difficult to 
fathom'': The spin-component of an electron has two possible values, yet it lacks an actual 
value; the neutron traverses the interferometer, yet it does not pass through a particular 
arm of the interferometer. Given the extrinsic nature of the matrix variables, the meaning of 
these statements ought to be clear. If something indicates the value of a particular spin 
component, it indicates either of two possible values, and the indicated value is possessed 
at the time of indication; if nothing indicates a value for a given time, no value is 
possessed at that time. By the same token, if the neutron's passage through the 
interferometer can be inferred from the facts then the neutron did traverse the 
interferometer, and if nothing indicates the arm taken by the neutron then the neutron did 
not pass through a particular arm. If this last statement is still difficult to fathom, it is 
because we approach the quantum world with an inadequate concept of space 
(Secs.~\ref{PSVPS1} and \ref{PSVPS2}).

The authors for their part make statements that others may find difficult to fathom: ``a 
matrix variable manifests itself on the spacetime scene, without entering this scene''; ``the 
clicks can be classified as electron clicks, neutron clicks, etc., although there are no 
electrons and neutrons on the spacetime scene''; ``[s]omething has happened to the 
source'' yet ``no events have taken place in the source.''

These statements, too, cease to be obscure when the extrinsic nature of matrix variables is 
taken into account. The authors make a sharp distinction between the matrix variables of 
the theory and what takes place in spacetime or is present on the spacetime scene. On the 
spacetime scene we have events such as the clicks, involving material objects such as 
counters, and ``all physical phenomena including clicks are described in terms of variables 
that each has a value, at any time.'' A matrix variable, on the other hand, does not have a 
value under any circumstance.'' It never enters the spacetime scene, but it ``can manifest 
itself (appear) with a value.''

The reason why this distinction is justified is that a 
fundamental physical theory which is essentially a probability algorithm presupposes 
(i)~actual events to which probabilities can be assigned and (ii)~actual events or states of 
affairs on the basis of which probabilities can be assigned. These events or states of affairs 
are described in terms of {\it intrinsic\/} variables---variables that possess values at all 
times. That a matrix variable never has a value and never enters the spacetime scene is a 
way of saying that it is not an intrinsic variable: It never has a value {\it by itself\/}. 
Correspondingly, that a matrix variable can manifest itself on the spacetime scene with 
this or that value is a way of saying that it is an extrinsic variable: It has a value only if, and to the extent that, a value is indicated.

Suppose that we perform a series of position measurements, and that every measurement 
yields exactly one result (that is, each time exactly one detector clicks). Then we are 
entitled to infer the existence of a persistent object, to think of the clicks given off by the 
detectors as indicating the successive positions of this object, to think of the behavior of 
the detectors as position measurements, and to think of the detectors as detectors. If 
instead each time exactly two detectors click (including the possibility that the same 
detector clicks twice), we are entitled to infer the existence of two objects (but not of two 
distinct individuals, unless distinguishing properties are also indicated). The upshot is 
that the number of ``components'' of a quantum system is as much an extrinsic variable of 
the system as the positions of its components. 

Thus when Ulfbeck and Bohr state that there are no electrons and neutrons on the 
spacetime scene, I take it to mean that the number of electrons or neutrons present in a 
system is not an intrinsic variable: It never has a value by itself. It is not the case that 
electron clicks happen because there are self-existent electrons that occasionally enter 
electron counters and produce clicks; rather, electrons exist only because each electron 
click indicates the presence of an electron in a particular region at a particular time. The 
electron's presence supervenes on the click. (``Supervenience'' is a suitable philosophical 
term for the relation between extrinsic values and value-indicating facts. The extrinsic values 
supervene on the value-indicating facts.)

When Ulfbeck and Bohr state that the source (``defined in terms of a distribution of clicks 
observed, for example, at the boundary of the source'') ``constitutes an object in space, 
with non-classical properties that can be established by the observation of distributions of 
clicks,'' the properties they refer to are non-classical in that they are extrinsic. At first sight 
the authors' statements that ``[s]omething has happened to the source'' yet ``no 
events have taken place in the source'' seem inconsistent. If due account is taken of the 
extrinsic nature of those properties, the apparent inconsistency disappears. As 
the authors themselves explain, ``the events that tell us that something has happened to 
the source\dots do not take place in the source.'' In other words, what happened 
to the source supervenes on a click that occurs at 
the boundary of the source. While it is correct that ``nothing takes place in the 
source that could be a cause of the click in the counter,'' saying that ``no events have taken 
place in the source'' is correct only in the sense that no intrinsic events have taken 
place there---no event has occurred by itself and triggered the counter. This does not rule 
out events in the source that supervene on the clicks at the boundary of the source.

\section{\large EXPERIENCE AND REALITY}

My third objection concerns the reality claims that go with the authors' distinction 
between matrix variables and what is present or takes place on the spacetime scene. 
Central to their conceptual scheme is (i)~``the identification of reality with the world of 
experience'' and (ii)~``the identification of experience with what takes place on the 
spacetime scene.'' From these identifications it follows that a matrix variable, which 
``cannot enter the spacetime 
scene,'' lacks reality. Ulfbeck and Bohr confirm this when they explain why they avoid 
expressions like ``the world of matrix variables'': ``it might convey the notion that 
something exists beyond the world of experience.'' At the same time they affirm that ``the 
matrix variables are physical quantities,'' and that they are ``variables in their own right.'' 
These are conflicting claims. How can a physical quantity that is a variable in its own right 
not be real?

The authors state that ``the locality permeating the quantal formalism is a symbolic one 
(which can be expressed in terms of fields that are associated with spacetime points but 
are not themselves on the spacetime scene since they derive from matrix variables).'' What 
can be associated with a spacetime point without being present at that point? The obvious 
answer: the probability for something to happen or be present at that point. Denying 
reality to a matrix variable comes down to saying that the probability for something to 
happen or be present in a given region~$R$ at a given time~$t$ is not something that 
exists either in $R$ or at~$t$. A matrix variable is unreal for the same reason that a 
possibility {\it per se\/} is not an actuality, and it can manifest itself on the spacetime scene 
for the same reason that something that is possible can be actual as well. There is no need 
to support these truisms with (inconsistent) reality claims, let alone to support these 
reality claims with metaphysical claims about experience and the spacetime scene.

By definition, empirical science deals with the world of experience. For this reason it 
cannot possibly tell us anything about what lies beyond the world of experience. Nor is 
experience (in the broadest sense of the term) amenable to empirical investigation since 
experience (in this sense) encompasses the world of experience and therefore does not take 
place within it. By confining reality to ``the world of experience'' Ulfbeck and Bohr make 
it clear that they refer to experience in this broadest sense, and this makes their reference 
to ``experience'' gratuitous and irrelevant.

Experience becomes accessible to scientific investigation in the attenuated sense of a 
relation between one part of the world of experience and another. Psychologists and 
neuroscientists have arrived at an impressive understanding of the relation between the 
``external'' world---one part of the world of experience---and its representation in the 
mind or by the brain---another part of the world of experience,---even though the ``final'' 
step to subjectivity remains shrouded in mystery since it concerns the existence of 
experience in its broadest sense, rather than the relation between one part of the world of 
experience and another. There is thus considerable {\it a posteriori\/} evidence---at least 
within psychology---of the usefulness of the division of the world of experience into a 
physical part---the world ``out there'' including the scientifically known brain---and a 
phenomenal part---the world as we perceive it.

I believe that the distinction between the phenomenal world---the world as we perceive 
it---and the physical world---the world as we theoretically conceive it---is equally useful 
for elucidating the ontological import of QM. Niels Bohr wrote:
\bq
It is my personal opinion that these [interpretational] difficulties are of such a nature that 
they hardly allow us to hope that we shall be able, within the world of the atom, to carry 
through a description in space and time that corresponds to our ordinary sensory 
perceptions.~\cite{Bohr23}
\eq
Bohr did not say that a spatiotemporal description was impossible but only that such a 
description could not be modeled after our ordinary sensory perceptions. The distinction 
between the physical and phenomenal worlds allows us to distinguish between the 
spatiotemporal features of the phenomenal world, which correspond to our ordinary 
sensory perceptions, and the spatiotemporal features of the physical world, which don't. 
And it may well be that the key to understanding QM lies in the unfamiliar 
spatiotemporal features of the quantum world---the physical world as described by QM. 
If so, Ulfbeck and Bohr could be blamed for delaying the progress of science just as 
Kant has been blamed for having delayed the discovery of non-Euclidean 
geometries by arguing that the validity of Euclidean geometry was {\it a priori\/} 
certain~\cite{KacUlam}.

The milestones in the history of science can be characterized by the recognition that what 
was once unquestioningly accepted---e.g., Euclidean geometry, absolute 
simul\-ta\-neity---actually lies within the compass of empirical falsifiabily. Ulfbeck and Bohr 
adhere to the Kantian doctrine that ``[s]pace and time constitute a scene established for 
the ordering of experiences,'' according to which spatiotemporal concepts have 
meaningful application only to what appears on the mental canvas woven out of space and 
time. In so doing they arbitrarily confine empirical reality to what is accessible to direct 
sensory experience, and render scientifically unassailable concepts of space and time that 
corresponds to our ordinary sensory perceptions. Thus, like Kant, they could be accused 
of delaying an adequate understanding of the spatiotemporal aspects of the physical 
world, and hence of QM.

\section{\large PHYSICAL SPACE VS PHENOMENAL SPACE---I}
\label{PSVPS1}
Making sense of QM is not so much a question about the ontological status of density 
operators---they are just sophisticated probability measures---as a question about the 
ontological status of {\it the space and time coordinates\/} that appear as arguments of 
density operators in the position representation. The demonstration that these coordinates 
cannot refer to the self-existent and intrinsically differentiated spatiotemporal background 
of classical physics requires nothing more elaborate than a two-slit experiment with 
electrons~\cite{Feynmanetal65}.

No electron is detected in the absence of the electron source in front of the slit plate, and 
no electron is detected behind the slit plate whenever the two slits are closed. This 
warrants the inference that each detected electron went through $L\&R$, the regions 
defined by the slits considered as one region. At the same time the existence of 
interference fringes implies that each electron went through $L\&R$ without going 
through a particular slit and, of course, without having been split into parts that went 
through different slits. (Like Ulfbeck and Bohr, I am concerned with the interpretation of 
standard QM unadulterated with, e.g., Bohmian trajectories~\cite{Bohm} or nonlinear 
modifications of the ``dynamics''~\cite{GRW}.) But if space were something that existed 
by itself, independently of its material ``content,'' and if it were made up of distinct, 
separate regions, every material object would be affected by this. No material object could 
be present in $L\&R$ without either being wholly contained in one of the regions or 
having a part in each region.

Interference fringes have been observed using C$_{60}$ molecules and a grating with 
50-nm-wide slits and a 100-nm period~\cite{Arndtetal}. Do we need any further proof that 
(in the context of standard QM) $L$ and $R$ cannot be distinct, self-existent ``parts of 
space,'' and that, consequently, space cannot be a self-existent and intrinsically 
partitioned expanse?

Although we readily agree that red, or a smile, cannot exist without a red object or a 
smiling face, we just as readily believe that positions can exist without being properties of 
material objects. We are prepared to think of material objects as substances, and we are 
not prepared to think of their properties as substances---except for their positions. (A 
substance is anything that can exist without being the property of something else.) There 
are reasons for these disparate attitudes, but they are psychological and neurobiological. 
They concern the co-production, by the mind and the brain, of the phenomenal world. 
They do not apply to the quantum world, but they certainly make it hard to make sense of 
it~\cite{BCCP,QMCCP}.

What is ultimately responsible for the disparate ways in which we handle positional 
information and other sensory data is the process by which the mind/brain 
integrates into phenomenal objects such phenomenal variables as hue, lightness, shape, 
and motion. This integration is based on positional information~\cite{Clark}. Phenomenal 
variables that occur in the same place are perceived as features of the same object, despite 
being neurally represented in separate feature maps. (A feature map is a layer of the 
neocortex in which cells map a particular phenomenal variable in such a way that adjacent 
cells generally correspond to adjacent locations in the visual field. In the macaque monkey 
as many as 32 distinct visual feature maps have been identified.) Thus while every 
phenomenal variable except location has at least one separate map, locations are present 
in all maps as the integrating factors. This unique role played by positional information in 
the process of feature integration is one of the reasons why we conceive of space as a 
pre-existent and intrinsically differentiated expanse. The neural process of feature 
integration, on which the creation of the phenomenal world is based, can only work if 
distinct locations somehow pre-exist. The creation of a {\it physical\/} object, on the other 
hand, is not a process that involves the integration of perceived features, so there is 
no reason to believe that physical space pre-exists as an intrinsically differentiated 
expanse. In fact, as the above analysis of an interference experiment has shown, this belief 
is inconsistent with~QM.

In the physical world the so-called ``parts of space'' are not real and distinct {\it per se\/}. 
A spatial region has a {\it contingent\/} reality, in the sense that it may exist for one 
material object and not exist for another. This calls for a criterion, and the obvious 
criterion is this: A region $V$ is real for an object $O$ if and only if the proposition 
``$O$~is in~$V$''---symbolically, $O{\rightarrow}V$---has a truth value. And the 
sufficient and necessary condition for the existence of a truth value is that one is indicated. 
But the truth or falsity of $O{\rightarrow}V$---can be indicated only if~$V$ exists, and 
for this is must be realized (made real) by being a possessed position that can be 
consistently considered intrinsic. This takes us to the genuine core of the so-called 
measurement problem, whose solution requires a demonstration of the consistent 
coexistence of extrinsic and intrinsic variables.

\section{\large SOLUTION OF THE MEASUREMENT PROBLEM}

There are objects whose indicated positions are so correlated that every one of them is 
consistent with every prediction that is based on previous indicated positions and a 
classical law of motion (except, of course, when the indicated positions serve to indicate 
unpredictable values). If I take this characterization as a definition of ``macroscopic 
object,'' I need to show that such objects exist. Note that this definition does not require 
that the probability of finding a macroscopic object where classically it could not be, is 
strictly~0. What it requires is that there be no position-indicating fact that is inconsistent 
with predictions based on a classical law of motion and earlier position-indicating facts.

The departure of an object $O$ from a classical trajectory can be indicated only if there 
are detectors whose position probability distributions are narrower than~$O$'s. Such 
detectors do not exist for all objects. Some objects have the sharpest positions in 
existence. For these objects the probability of a position-indicating event that is 
inconsistent with a classical trajectory is necessarily very low. It is therefore certain that 
{\it among\/} these objects there will be macroscopic ones.

Since no object has an exact position, it might be argued that even for a macroscopic 
object~$M$ there always exists a small enough region~$V$ such that the proposition 
$M{\rightarrow}V$ lacks a truth value. But this is an error. Macroscopic objects have the 
sharpest positions in existence. There isn't any object that has a sharper 
position. {\it A fortiori\/}, there isn't any object for which $V$ is real. But a region exists 
only if it is real for at least one material object. It follows that there exists no region $V$ 
such that the proposition $M{\rightarrow}V$ lacks a truth value. Such a region may exist 
in our imagination, but it does not exist in the real world.

Now recall why positions are extrinsic: The proposition $O{\rightarrow}V$ may or may 
not have a truth value. One therefore needs a criterion for the existence of a truth value: A 
truth value must be indicated. But one doesn't need a criterion for the existence of a truth 
value if for every {\it existing\/} region~$V$ the proposition $M{\rightarrow}V$ {\it has\/} 
a truth value. Since macroscopic objects satisfy this condition, their positions can be 
consistently considered intrinsic. We can think of the positions of macroscopic objects 
(macroscopic positions, for short) as forming a system of causally connected properties 
that are effectively (that is, for all quantitative purposes rather than merely all practical 
ones) detached from the facts by which they are indicated. We can think of this system as 
a self-existent and self-contained causal nexus interspersed with transitions (of 
value-indicating positions) that are causally linked to the future but not to the past.

Since the beginning of time (in about 1926) it has been argued that QM is about 
experience, knowledge, or information~\cite{LonBau,vN,Wigner,Heisbg58}, rather than 
about a free-standing reality capable of being described without reference to observers, 
their information, their interventions into ``the course of Nature''~\cite{FuPer}, or their 
arbitrary decisions as to where to make the ``shifty split'' between ``system'' and 
``apparatus''~\cite{Bell}. Why? Because it is such an easy way to establish the consistent 
coexistence of extrinsic and intrinsic variables. If the properties of the quantum world are 
extrinsic (that is, if they ``dangle'' from, or supervene on, something), and if the quantum 
world is coextensive with the physical world, then from what can they ``dangle''? The 
obvious answer: from us, from what we perceive, or from what we know.

For this easy way out we pay a high price. By safeguarding against empirical refutation 
conceptions of space and time that are consistent with the phenomenal world but 
inconsistent with the physical world, we make sure that we won't discover the 
spatiotemporal features of the quantum world. And by rooting the possible 
value-indicating events, to which QM assigns probabilities, in the world of sensory 
experience, we make sure that we can't conceive of the quantum world as a strongly 
objective, free-standing reality that owes nothing to observers, information, or our 
interventions into the course of Nature.

\section{\large PHYSICAL SPACE VS PHENOMENAL SPACE---II}
\label{PSVPS2}
Macroscopic positions are so abundantly and so sharply indicated that they are only 
counterfactually fuzzy. Their fuzziness never evinces itself, through uncaused transitions 
or in any other manner. It exists solely in relation to an imaginary spatial background that 
is more differentiated than the physical world. The space over which the position of a 
macroscopic object is ``smeared out'' is never probed. This space is undifferentiated; it 
contains no smaller regions. We may imagine smaller regions, but they have no 
counterparts in the physical world. The distinctions we make between them are 
distinctions that nature does not make.

It follows that the quantum world is only finitely differentiated spacewise, and that it 
ought to regarded as constructed {\it from the top down\/}, by a finite process of 
differentiation, rather than from the bottom up, on a self-existent and maximally 
differentiated spatial expanse. And much the same applies to the world's temporal aspect. 
Time is not an independent observable; it has to be read off of deterministically evolving 
positions---the positions of macroscopic clocks. If these bear a residual fuzziness, so do all 
indicated times. The upshot: The quantum world is maximally differentiated neither 
spacewise nor timewise, and it is constructed from the top down with respect to both space 
and time.

To advance further, we must be clear about what it means when a particle is said to be 
``pointlike.'' This is an expression of the fact that the particle lacks internal structure. 
Nothing in the formalism of QM refers to the shape of an object that lacks internal 
structure, and the empirical data cannot possibly do so. All that experiments can reveal in 
this regard is the absence of evidence of internal structure. The idea that a so-called 
``point particle'' is an object that not only lacks internal relations but also has the shape of 
a point, is thus unwarranted both theoretically and experimentally. It is, besides, seriously 
misleading, inasmuch as the image of a pointlike object suggests the existence of an 
infinitesimal neighborhood in an intrinsically and maximally differentiated spatial expanse. To 
bring our intuitions in line with the spatiotemporal aspects of the quantum world, we 
need to conceive of all so-called ``point particles'' as {\it formless\/} objects. What lacks 
internal relations also lacks a shape.

It follows that the shapes of material objects resolve themselves into sets of (more or less 
fuzzy) spatial relations between formless objects, and that space itself is the totality of 
such relations---relative positions and relative orientations. It further follows that the 
corresponding relata {\it do not exist in space\/}. Space contains, in the proper, 
set-theoretic sense of ``containment,'' the forms of all things that have forms---for forms 
are sets of spatial relations---but it does not contain material objects over and above their 
forms; {\it a fortiori\/} it does not contain the formless constituents of matter. Instead, 
space exists {\it between\/} them; it is spanned by their relations.

The quantum world with its fuzzy spatial relations does not ``fit'' into the self-existent and 
maximally differentiated expanse of classical space; the possibility of thinking of the relata 
as points and embedding them in a single manifold exists only if all spatial relations 
are definite (``sharp''). A clear distinction should therefore be made between the existing 
(more or less fuzzy) spatial relations that constitute physical space, and the purely imaginary 
space that comes with each material object~$O$ and contains the unpossessed exact 
positions relative to~$O$. These imaginary spaces are delocalized relative to each other: The 
unpossessed exact positions relative to~$O$ are fuzzy relative to any material object other 
than~$O$.

The difference between the respective ways in which spatial distinctions are realized in the 
physical and phenomenal worlds could hardly be greater. In the physical world spatial 
distinctions are realized by means of (more or less fuzzy) spatial relations between 
formless objects. In the phenomenal world they are realized by means of boundaries. 
Visual representations arise by way of an analysis of the visual field that capitalizes on 
contrast information. Data arriving at the visual cortex from homogeneously colored and 
evenly lit regions of 
the visual field do not make it into conscious awareness. Such regions are filled in on the 
basis of contrast information across their boundaries~\cite{QMCCP}. (This explains, 
among many other things, why the blind spot goes unperceived whenever it falls in such a 
region.) The way in which the brain processes visual information thus guarantees that the 
result---the phenomenal world---is a world of objects whose shapes are bounding 
surfaces. The parts of any {\it phenomenal\/} object accordingly are defined by the parts 
of the space it ``occupies,'' and these are defined by delimiting and separating surfaces. 
This too implies that the parts of space pre-exist somehow---otherwise they couldn't 
define the parts of a phenomenal object,---and this is another reason why we tend to 
conceive of space (inconsistently with the quantum world) as a pre-existent and 
intrinsically differentiated expanse.

\section{\large SUBSTANCE AND THE QUANTUM WORLD}

There are other respects in which the physical world is built top-down rather than 
bottom-up. If all substances were intrinsically distinct, the world could be thought of as 
constructed from the bottom up, by aggregation. The same would be true in a 
deterministic world, since determinism allows us to associate a distinct substance 
with each possessed position.

For centuries philosophers have argued over the existence of intrinsically distinct 
substances. QM has settled the question for good: There are no intrinsically distinct 
substances. The concept of substance betokens {\it existence\/}; it {\it never\/} betokens 
individuality. Individuality is strictly a matter of properties. Given the extrinsic nature of 
properties, this means that a quantum system can be associated with distinct substances 
only to the extent that distinguishing properties are indicated. Since this extent is limited, 
the possibility of decomposing the quantum world into distinct substances is limited as 
well. Hence the quantum world ought to be regarded as constructed from the top down 
not only with regard to its spatial and temporal aspects but also with regard to its 
substantial aspect.

If we believe that QM is about regularities in sensory experience, we don't need the concept 
of ``substance.'' But we need it if we want to think of the quantum world as a 
free-standing reality, inasmuch as it is the concept of ``substance'' that betokens 
independent existence. And we need to know how the quantum world relates to its 
substance. Since it would be absurd to substantialize a probability algorithm, 
substantiality can't be attached to a state vector or a wave function, as Ulfbeck and Bohr 
correctly point out. Nor can it be attached to the points of a spacetime manifold, as the 
previous sections have shown. Nor can the substance of the quantum world be 
decomposed into a multiplicity of intrinsically distinct substances, as we just saw.

If the property of being here and the property of being there are simultaneously 
possessed, how many substances does that make? The correct answer is ``one,'' for the 
substance that betokens the reality of the property of being here also betokens the reality 
of the property of being there. QM does not permit us to interpose a multiplicity of 
distinct substances between the substance that betokens existence and the multiplicity of 
possessed positions. QM thus lends unstinting support to the constitutive idea of all 
monistic ontologies: Ultimately there is only one substance. As physicists we are not 
concerned with the intrinsic nature of this substance. (It arguably plays an important role 
in the emergence of consciousness). What is of interest to us is how it acquires the aspect 
of a spatiotemporal expanse teeming with quarks and leptons.

In broad outline the answer is simple enough: By entering into spatial relations with itself, 
this substance acquires at one stroke the aspect of a multiplicity of spatial relations, which 
constitute forms and space, and the aspect of a multiplicity of formless relata, which 
constitute matter. And if we allow the spatial relations to change, we have time as well, for 
change and time are co-implicates. (In a timeless world nothing can change, and a world 
in which nothing changes is a world without temporal relations; such a world is 
temporally undifferentiated and therefore timeless, just as a world without spatial 
relations is spatially undifferentiated and therefore spaceless.)

\section{\large WHAT HAPPENS BETWEEN VALUE-INDICATING\\
FACTS?}

What can be said about the interval between two times for which properties are indicated 
if no property is indicated for any intermediate time? To begin with, consider again a two-slit 
experiment in which all that is indicated is each electron's place of departure in 
front of the slit plate and its place of arrival behind the slit plate. While the 
propositions $e{\rightarrow}L$ (``the electron went through $L$'') and 
$e{\rightarrow}R$ lack truth values, the corresponding histories contribute to the 
observed probability distribution. We therefore have reason to conclude that both 
histories happen, but {\it indistinguishably\/}, in the sense that the distinction we make 
between them is a distinction that nature does not make. The reason why we cannot assign 
separate truth values to $e{\rightarrow}L$ and $e{\rightarrow}R$ but only a single truth 
value to $e{\rightarrow}L\&R$ is that the conceptual difference between the two histories 
has no counterpart in the physical world.

The probability of detecting at a given time and location a particle having last been 
``seen'' at another time and location, is determined by a ``propagator'' 
$K(x_2,t_2;x_1,t_1)$ that can be calculated by summing over all continuous paths leading 
from $(x_1,t_1)$ to $(x_2,t_2)$~\cite{Feynman48,FH65}. Hence nothing stands in the 
way of the claim that if the particle is present at these two spacetime locations, and if 
nothing indicates its intermediate whereabouts, then it does travel along all of those paths, 
subject to the understanding that the distinctions we make between these histories 
correspond to nothing in the physical world.

If the initial and final states of affairs include indistinguishable particles, there are further 
distinctions that nature does not make. The distinctions we make between histories that 
connect particular incoming particles with particular outgoing particles of the same type, 
which are based on the false notion that particles are intrinsically distinct substances, also 
correspond to nothing in the physical world. Such histories happen together, 
indistinguishably, in the sense 
spelled out above. Finally, if the possibility of pair events is taken into account, the 
histories that happen together, indistinguishably, are constrained only by conservation 
laws. In general it can be said that whenever QM requires the addition of amplitudes, the 
distinction we make between the corresponding histories do not exist in the physical 
world.

It is obvious that this interpretation crucially depends on the extrinsic nature of quantum 
variables. It is clearly impossible to construe a sum over histories leading from one state 
of affairs to another as something that happens ``by itself,'' rather than as something that 
supervenes on these states of affairs. It is not the case that an electron is detected by a 
detector {\it because\/} it followed all continuous paths leading from the source to the 
detector. Rather, the electron followed all these paths because it left the source and was 
detected by the detector, and because there are no matters of fact about its intermediate 
whereabouts.

Every value-indicating fact realizes the difference between a range of possible values, 
inasmuch as what is indicated is not only the possession of a particular value (the truth of 
one proposition) but also the non-possession of all other possible values (the falsity of $n{-
}1$ propositions, $n$ being the number of possible values). (Recall that our distinctions 
between alternatives correspond to something in the physical world if and only if truth values 
are indicated for the corresponding propositions.) Every value-indicating fact therefore 
reduces the set of 
histories that happen indistinguishably. Hence in a sense the world is built top-down also 
dynamically: from ``everything at once'' to something specific, by an elimination of 
alternatives. Every value-indicating fact reduces the set of histories that happen together, 
yet even the totality of value-indicating facts does not reduce it to a single history.

\section{\large CONCLUSION}

The metaphysical presuppositions of Ulfbeck and Bohr effectively safeguard against 
empirical refutation conceptions of space and time that are essentially classical. Claims by 
these authors to the effect that genuine fortuitousness ``does not invoke notions taken over 
from classical physics,'' and that matrix variables are ``entirely liberated'' from such 
notions, must therefore be taken with a grain of salt.

An alternative view of QM has been presented, which is liberated from such notions to an 
extent that permits conceiving of the quantum world as a free-standing reality. Central to 
this view is a conceptual clarification of the indefiniteness that is crucial for the stability of 
spatially extended material objects. This requires novel conceptions of space and time, 
and it implies a top-down structure for the quantum world: Rather than being built 
bottom-up, on a maximally differentiated spatiotemporal manifold, the quantum world 
arises from a limited spatiotemporal differentiation. A similar top-down structure 
characterizes its substantial and dynamical aspects.

\end{document}